\documentclass{aa}
\usepackage{graphicx}
\usepackage{natbib}
\def\bp{\object{$\beta$\,Pictoris}}
\def\mic{\object{AU\,Mic}}
\newcommand{\ma}[1]{\mathrm{#1}}

\newcommand{\dma}[1]{_{\mathrm{#1}}}
\def\d{\mathrm d}
\begin{document}
\title{Upper limit on the gas density in the \bp\ system}
\subtitle{On the effect of gas drag on the dust dynamics}

\author{P. Th\'ebault\inst{1,2}, J.-C. Augereau\inst{3,4}}
\institute{
Stockholm Observatory, Albanova Universitetcentrum, SE-10691 Stockholm,
Sweden
\and
Observatoire de Paris, Section de Meudon,
F-92195 Meudon Principal Cedex, France
\and 
Leiden Observatory, PO Box 9513, 2300 RA Leiden, The Netherlands
\and
Laboratoire d'Astrophysique de l'Observatoire de Grenoble, B.P. 53,
38041 Grenoble Cedex 9, France}

\offprints{P. Th\'ebault} \mail{philippe.thebault@obspm.fr}
\date{Received; accepted} \titlerunning{Gas density in \bp}
\authorrunning{Th\'ebault and Augereau}

\abstract{
We investigate in this paper the effect of gas drag on the dynamics
of the dust particles in the edge-on \bp\ disc in order to derive an
upper limit on the mass of gas
in this system. Our study is motivated by the large uncertainties 
on the amount of gas in the \bp\ disc currently found in the literature.
The dust particles are assumed to originate from a colliding annulus of planetesimals
peaked around 100\,AU from the central star as proposed by Augereau et al.\,(2001).
We consider the various gas densities that have been inferred from independent
observing techniques and we discuss their impact on the dust dynamics and 
on the disc profile in scattered light along the midplane. We show that
the observed scattered light profile of the disc cannot be properly
reproduced if the hygrogen gas number density at
117\,AU exceeds $10^4$\,cm$^{-3}$.
This corresponds to an upper limit on the total gas mass of about
$0.4\,$M$_{\oplus}$ assuming the gas density profile inferred
by Brandeker et al.\,(2004) and thus to a gas to dust mass ratio
smaller than 1.
Our approach therefore provides an independent diagnostic for a
gas depletion in the \bp\ system relative to the dust disc.
Such an approach could also be used to constrain the gas content of
recently identified systems like the edge-on disc around \mic.

\keywords{stars: planetary systems -- stars: \bp
        -- planetary systems: formation --
        planets and satellites: formation
               } }
\maketitle

\section{Introduction}

The \bp\ system is still one the best known examples of a
debris disc surrounding a young Main Sequence star.
The dust component of the almost edge-on disc has been 
thoroughly observationally investigated, in thermal emission
as well as in scattered light \citep[e.g.][and references therein]{arty97}.
The total dust mass is believed to range between a few lunar masses
and $0.5\,M_{\oplus}$ \citep[e.g.][]{zuc93,arty97,den00}. These estimates
lead to collisional lifetimes of dust particles which are much shorter than
the estimated age of the system \citep[$8$--$20$\,Myr,][]{zuc01,song03,dif04},
which implies that the dust disc cannot be primordial but
has to be replenished by
collisions or evaporation of larger bodies \citep[e.g.][]{arty97,lec96}.
Scattered light midplane surface brightness profiles display a sharp
decrease beyond 120\,AU \citep{kal95} where the luminosity profile
follows a steep radial power law $r^{\alpha}$ 
with $-5.5<\alpha<-4.5$ \citep{heap00}. These outer regions
of the disc that extend over hundreds of AU are believed
to be made of small grains primarily produced
within $150\,$AU and placed on high eccentricity orbits
by radiation pressure \citep{lec96,aug01}.

In contrast to our relatively good knowledge of the
dust distribution, the gaseous compound is still poorly constrained.
This is in particular the case for hydrogen, which should {\it a priori}
make up most of the gas disc. So far, attempts to directly
detect hydrogen have failed, with the exception of \citet{thi01} who
have measured with ISO H$_2$ infrared emission features associated
to pure rotational transitions. The mid-infrared observations
are consistent with more than
$50\,M_{\oplus}$ of warm H$_{2}$ in the \bp\ disc.
These results are challenged by the non-detection in
the UV with the FUSE satellite of absorption lines from H$_{2}$
electronic transition \citep{lec01}. This non-detection places
an upper limit on the H$_{2}$ column density three order of magnitude
smaller than the value derived from the infrared observations.

The H$_2$ emission is unlikely to be of interstellar origin according
to \citet{lec96}. The ISO aperture been large, the infrared emission could
originate from regions not probed by FUSE if the gas disc is very
spatially extended.
Recent mid-infrared observations with Spitzer nevertheless
do no confirm the 17.035$\,\mu$m H$_2$ emission feature observed with
ISO \citep{che04}. The gas mass upper limit based on the Spitzer results
is 11\,M$_{\oplus}$ assuming a gas temperature of 110\,K.

Attempts have also been made to indirectly reconstruct the hydrogen abundance
from other more accurately observed species.
Absorption lines in the stellar spectrum have indeed revealed
the presence of numerous metallic elements \citep[e.g.][]
{hobbs85,lag98,rob00}. From early estimates, rescaling to solar
abundances lead to column densities of
$N({\rm H) = 10}^{18}$--$4\,10^{20}{\rm cm}^{-2}$
\citep{hobbs85}. \citet{kamp03} performed
emission line calculations for different gas tracers (CO but also
C and C$^{+}$) in the disc. They conclude that
the total \bp\ gas mass (presumably mostly made of hydrogen)
would lie between 0.2 and 4\,$M_{\oplus}$.
More recently, \citet{brand04}, using the VLT/UVES instrument,
spatially resolved the gas disc and observed numerous emission lines,
mainly from FeI, CaII and NaI which was observed as far as 323\,AU from
the star. Estimating a ionisation structure for the disc
and assuming solar abundances, they found a column density
for atomic hydrogen of $N({\rm HI) = 8\,10}^{18}{\rm cm}^{-2}$, consistent
with the upper limit of \citet{freud95}, and 
$N({\rm H}_{2}) = 3\,10^{18}{\rm cm}^{-2}$ for molecular hydrogen,
consistent with the \citet{lec01} upper limit. However, the inferred
total mass of the gas disc, $\simeq 0.1\,M_{\oplus}$, is inconsistent
with the $50\,M_{\oplus}$ gas mass measured with ISO by \citet{thi01}.

Nevertheless, independent considerations might argue in favour of
a more massive gas disc. One puzzling result deduced from the observed
absorption lines is indeed that all ion species seem to be on keplerian
orbits, with low relative radial velocities \citep{olof01}.
This is in strong contradiction with the fact that many
of these elements should be rapidly ejected on unbound orbits
by strong stellar radiation pressure (\bp\ is an A5V star).
Several explanations have been proposed to explain this contradiction.
One of them is that gaseous friction due to an unseen gaseous braking
agent, possibly hydrogen, should damp the outward outflow
\citep{liseau03}. With calculations based on the \citet{lis99}
model, \citet{brand04} estimated that a high density gas disc
of $\simeq\,50\,M_{\oplus}$ could lead to radial velocities
compatible with the observed ones provided the disc is
metal depleted.

There are thus still large uncertainties regarding the amount
of gas in the \bp\ disc. We propose here to
address this issue by looking at the effect of gaseous friction
on the dynamics of the dust grains observed in scattered light.
This approach is used to constrain the gas density in the disc
and its radial distribution. We base our approach on the results of
\citet{aug01}, who have shown that the observed dust disc could be
successfully model if the grains are produced by
a distribution of parent bodies within $150\,$AU from the central
star and placed on orbits with large semi-major axis ($a$)
and high eccentricities ($e$) by radiation pressure.
The important point is that this satisfying fit of the
surface brightness profile was obtained for a {\it gas free}
system. 

Our main goal is here to estimate how gas drag might affect these
results, and in particular which gas disc densities and distributions
yield dust density profiles still compatible with observations and
which ones don't and might thus be ruled out.
Our approach is much in the spirit of the remarkable pioneering work
of \citet{lec96}, who explored the scattered light profile
for different parent bodies prescriptions and even included gas drag, but\
considered an academic gas disc prescription ($\rho_{g} \propto 1/r$) and
restricted their approach to $\beta<1$ particles (note that
the behaviour of $\beta>1$ particles was addressed by
\citet{lec98} in the case of BD+31643, a very different stellar environment).
This neglection of the  $\beta>1$ particles
might be too restrictive in the presence of efficient gas drag
which prevents high $\beta$ particles from leaving the system
as quickly as in a gas free disc (see section 2.2 and Fig.\,\ref{bprofs}).
Furthermore, at that time the radial power law index for the outer
midplane luminosity profile was believed to be in the $[-3.6,-4.3]$
range rather than in the $[-4.5,-5.5]$ range.

\section{Model}

\subsection{Equations of motion}

Our integrator is a classic $4^\ma{th}\,$order Runge-Kutta
which test runs (performing mutual comparisons with higher order
algorithms) have proved
to be of a satisfying accuracy for the present study.
The integrator takes into account the star's central potential,
the radiation pressure force and gaseous friction.
The radiation pressure on a grain of mass $m$ at a distance $r$
from the central star is expressed as a function of the gravitational
potential trough the usual $\beta$ parameter
\begin{equation}
\vec{F\dma{rad}}=-\beta \vec{F\dma{grav}} = \beta \frac {GMm}{r^{2}} \vec{r} \,\,\, .
\label{rad}
\end{equation}
assuming an optically thin disc in all directions.
Here $G$ is the gravitational constant and $M$ the mass
of the central star.
We will consider grains larger than $0.1\,\mu$m
radius, for which there is an inverse one to one relation between
$\beta$ and grain size $s$ \citep{arty88} and we assume
$\beta \simeq 0.5\times\left(5\,\mu{\rm m}/s \right)$
with $s$ expressed in $\mu$m.
For the gas drag force we follow
\citet{take01} and adopt the \citet{kwo} formalism
\begin{equation}
\vec{F\dma{gas}}= -\pi\,\rho\dma{g}\,s^{2}\,\left( v\dma{T}^{2} +
\Delta v^{2} \right)^{1/2} \vec{\Delta v}
\label{eps}
\end{equation}
where $\rho\dma{g}$ is the gas density, $v\dma{T}$ is the gas thermal velocity
and {\bf $\Delta v$} is the difference between the grain velocity {\bf $v$}
and the gas velocity {\bf $v\dma{g}$} expressed as a function of the
keplerian velocity $v\dma{k}$

\begin{equation}
v\dma{g}=v\dma{K} (1-\nu)^{2} \,\,\, .
\label{vg}
\end{equation}
The parameter $\nu$ depends on the gas
pressure gradient $\d P\dma{g}/\d r$ and reads
\begin{equation}
\nu=\frac{1}{r\Omega^{2}\rho\dma{g}} \frac{\d P\dma{g}}{\d r} \, .
\label{nu}
\end{equation}

\subsection{Dust disc}

\begin{figure} 
\includegraphics[angle=0,origin=br,width=0.95\columnwidth]{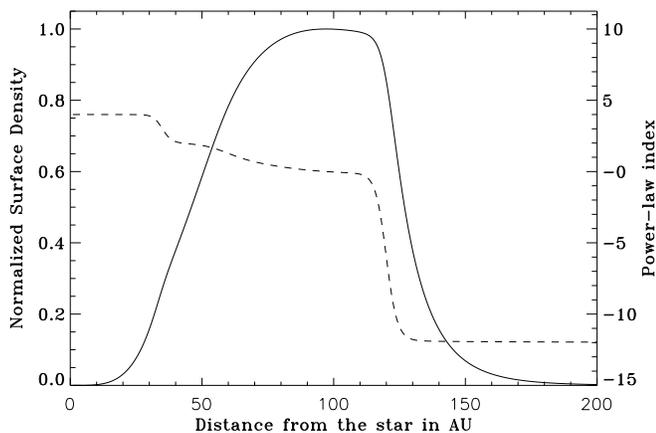} 
\caption[]{Surface density of the parent bodies disc responsible for
the resolved scattered light images according to 
\citet[][solid line]{aug01}. The dashed line
together with the right $y$ axis give the power-law index of the profile.} 
\label{surfdens} 
\end{figure}

We assume that the dust particles are produced by parent
bodies on circular orbits following the distribution derived by
\citet{aug01} and displayed in Figure\,\ref{surfdens}. This parent
body distribution has been shown to produce a spatially extended dust
disc that correctly matches the resolved scattered light images as well
as the long-wavelength photometric data. The parent body distribution
also results in a simulated disc radial brightness profile at
12\,$\mu$m consistent with the N-band thermal images obtained
by \citet{pan97} beyond $\sim$30\,AU.
But importantly, \citet{aug01} couldn't reproduce
the absolute flux that they attribute to an additional population
of very small grains with $\beta$ ratios likely larger than 0.5.
Recent mid-infrared observations support this picture \citep{oka04,tel05}.
The adopted distribution of parent bodies displayed in Fig.\,\ref{surfdens}
peaks in the 80-120\,AU region and cuts beyond
150\,AU (for the validity of this assumption in the context of
the present paper see discussion in section 4).

The differential size distribution of the produced dust is assumed
to follow the usual equilibrium power law $\d N = C_{0} s^{-3.5}\d s$
\citep{dohn69}. We then consider an arbitrary number
$N\dma{num}$ of particles satisfying this distribution
between $s\dma{min}=0.1\,\mu$m ($\beta=25$) and
$s\dma{max}=500\,\mu$m ($\beta=0.005$).
We consider objects up to high $\beta$ because gas drag might be very
efficient in slowing down the escape of these small grains, hence
increasing their contribution to the luminosity profile
(for a more quantitative analysis, see section 3).
In order to get statistical significance for all
size ranges despite the steep power law of the size distribution,
three different runs, each with $N\dma{num}=20000$, are performed for the
$0.005<\beta<0.05$, $0.05<\beta<0.5$ and $0.5<\beta<25$
populations, which are then recombined with the
appropriate weights.

At each moment every particle is assigned a collision
destruction probability, depending on its size, velocity and location
in the disc, and has thus a limited lifetime.
Due to obvious computational constraints, a realistic modeling
of collisions is here excluded. We use an empirical approach
where each particle is assigned, at each timestep, a collisional
destruction probability depending on its $\beta$ value, its distance $r$
from the central star and its radial velocity $v_{r}$

\begin{equation}
f\dma{coll} = \frac {\Delta t}{t\dma{coll}}
\label{fcoll}
\end{equation}
where $\Delta t$ is the simulation timestep
and $t\dma{coll}$  the collision time scale. We
approximate $t\dma{coll}$ by

\begin{equation}
t\dma{coll} = \left(\frac{\beta_{0}}{\beta}\right)^{\alpha}
\left(\frac{n_{100AU}}{n_{r}}\right) \left(\frac{r}{100{\rm AU}}\right)^{0.5}
\frac{v_{r_{0}}}{v_{r}}\, t\dma{coll_{0}}
\label{tcoll}
\end{equation}
where $t\dma{coll_{0}}\simeq 10^{4}$\,years is a reference collision
timescale that we take equal to the estimate
of \citet{arty97} for a $\beta_{0}=0.3$ particle at 100\,AU from the
central star. $v_{r_{0}}$ is
the radial velocity of the $\beta_{0}$ particles at 100\,AU and
$n_{R}$ is the volumic number dust density at distance $r$.
The $\beta$ ({\it i.e.} grain size) dependency of $t\dma{coll}$ 
relies on complex physical parameters. However, numerical studies
\citep[e.g. Fig.\,7 of \citet{wya02} and Fig.\,12 of][]{the03}
indicate that the coefficient $\alpha$ ranges between $0.3$ and $0.5$ for
small particles. We take here $\alpha=0.4$.
The absolute value of the collision
rates might be rescaled by changing the $t\dma{coll_{0}}$ parameter.

For $n_{r}$ in Eq.\,\ref{tcoll} we assume the dust density profile
derived by \citet{aug01} (see their Fig.\,2) from the parent bodies profile
displayed in Fig.\,\ref{surfdens}. A more coherent approach would have
consisted in taking into account $n_{r}$ matching the final dust
distribution for each individual simulation we performed.
But this would have added a lot of complexity to the problem by
introducing a circular argument, the final distribution being in
principle dependent on the assumed collision timescale.
We nevertheless empirically checked the validity of our
prescription by performing one test run with $n_{r}$ matching the final
dust distribution obtained in the reference case of Sec.\,3.1, and ended
up with a final luminosity profile which never departs by more than
$10\%$ from the nominal one. Possible errors due to the inevitable
departure between the assumed $n_{r}$ prescription and the final
dust distribution are thus expected to be negligible compared to the
impact of the parameters explored throughout the paper on the calculated
brightness profiles.

Such an empirical prescription is of course but a crude approximation
of the real collisional behaviour of dust particles. To the first
order, it provides nevertheless a good indication on how the collision rate
relies on the grain size, on the grain velocity and position in
the disc\footnote{a similar, though simpler,
cut-off lifetime prescription acting as a substitute for collisional dust
dectruction has been recently used by \citet{Del05}}.

\subsection{Gas disc characteristics}

Several test gas discs, with different profiles
and spatial extents, are explored.
Our reference case is the profile derived by
\citet{brand04} from their observations of metallic
emission lines, in particular NaI, out to 323\,AU.
From the observed NaI profile they deduced the
radial density profile of hydrogen nuclei to be
\begin{equation}
n({\rm H})=n_{0} \left[ \left(\frac{r}{r_{0}}\right)^{2.4} +
\left(\frac{r}{r_{0}}\right)^{5.3} \right]^{-\frac{1}{2}}
{\rm cm}^{-3}
\label{sod}
\end{equation}
with $r_0=117$\,AU. 

We will explore different $n_{0}$ values between the two extreme  values
considered by \citet{brand04}:
\begin{enumerate}
\item $n_{0}=2.25\,10^{3}$\,cm$^{-3}$,
required if solar abundances are assumed, which leads
to a total H mass of $\simeq\,0.1M_{\oplus}$,
\item a much higher density: $n_{0}=\,10^{6}$\,cm$^{-3}$
that implies a significant departure from solar abundances
but required, according to \citet{brand04}, if
hydrogen is to act as the "braking agent" maintaining all
high $\beta$ elements on keplerian orbits.
\end{enumerate}
We will also consider the case where the gas disc is truncated at 150\,AU. 
Finally, an academic case where the gas distribution matches the classical
\citet{haya81} radial profile proportional to $r^{-2.75}$ is also tested.
All initial conditions are summarized in Table\,\ref{tabinit}.

\begin{table*}[!tbph]
\caption{Summary of the different gas disc properties assumed in this paper.
The densities $n_0$ are given at a distance $r_0=117\,AU$ from the star.}
\label{tabinit}
\begin{tabular}{lcccl}
\hline
       & $n_0$\,[cm$^{-3}$] & radial profile
       & Total Mass\,[M$_{\oplus}$] 
       & comment \\
\hline
case 1 & $10^{6}$ & \citet{brand04} & $40\,M_{\oplus}$ & 
density required for the ``gaseous braking agent'' hypothesis\\
case 2 & $2.25\,10^{3}$ & \citet{brand04} & $0.1\,M_{\oplus}$ & 
reconstructed from the NaI profile assuming solar abundances\\
case 3 & $10^{4}$ & \citet{brand04} & $0.4\,M_{\oplus}$ &  \\
case 4 & $10^{5}$ & \citet{brand04} & $4\,M_{\oplus}$ &  \\
case 5 & $10^{6}$ & \citet{brand04} & $15\,M_{\oplus}$ & 
truncated at 150\,AU \\
case 6 & $7.10^{5}$ & Hayashi ($r^{-2.75}$) & $40\,M_{\oplus}$ &  \\
\hline
\end{tabular}
\end{table*}
\begin{figure*}
\makebox[\textwidth]{
\includegraphics[width=\columnwidth]{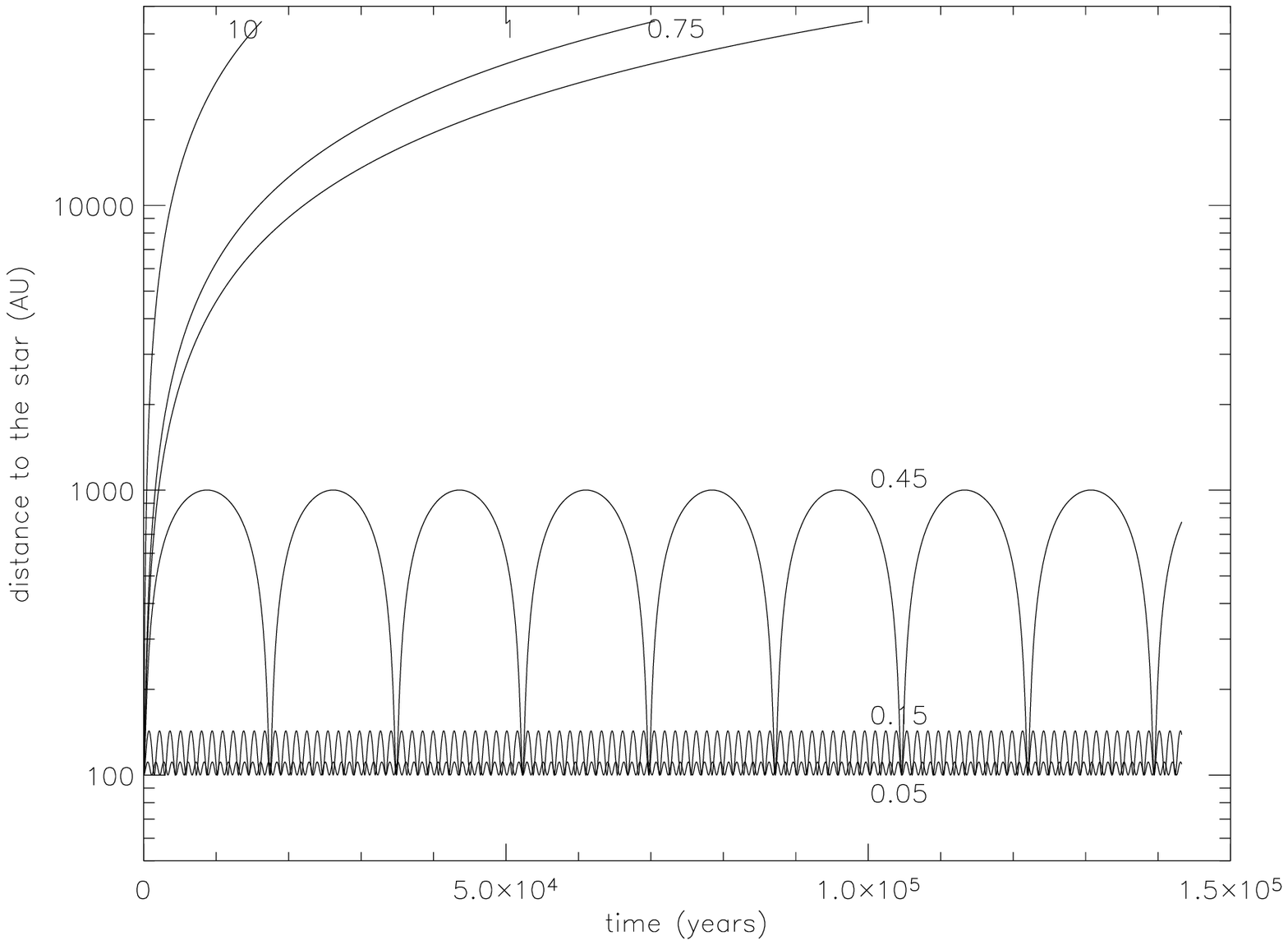}
\hfil
\includegraphics[width=\columnwidth]{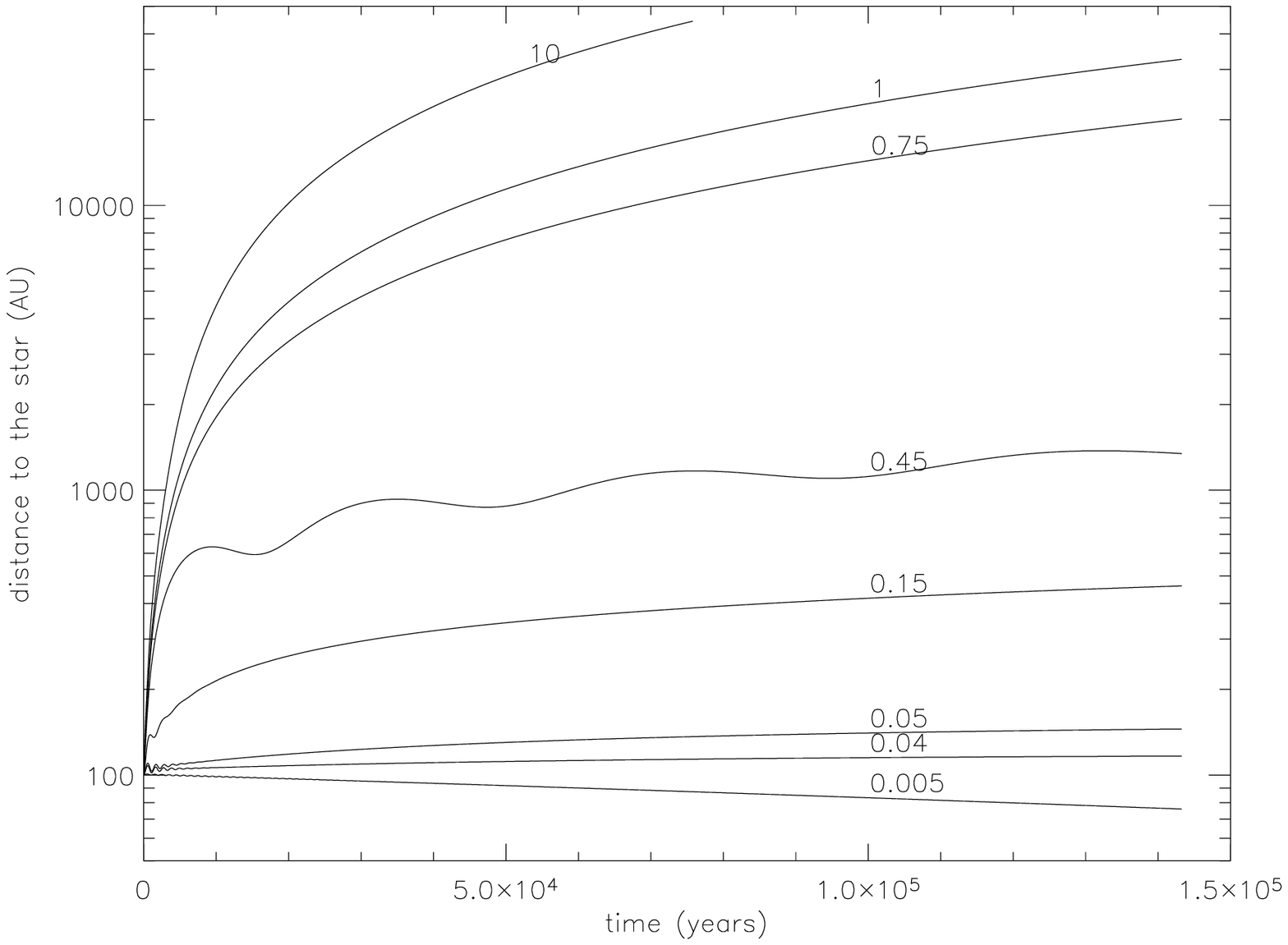}
}\\[-45ex]
\hspace*{0.4\columnwidth}
{\Large\textbf{a}}\hspace*{0.9\columnwidth}{\Large\textbf{b}}\\[43ex]
\caption[]{Radial position as a function of time, for different values
of $\beta$, for one particle initially located at 100 AU:
a) gas free case
b) high gas density case (case 1)
.
Note how efficient gas drag partially erases the abrupt transition at
$\beta=0.5$ for the dynamical behaviour of small grains.
}
\label{bprofs}
\end{figure*}

\subsection{Scattered light profile}

The aim of the paper is to use the shape of the scattered light
surface brightness profile along the disc midplane to set an upper
limit on the gas density in the \bp\ system. We detail in this
section the procedure we used to calculate the brightness profile.

All particles are produced at an initial moment $t_{0}$
after which we let the system dynamically evolve.
At {\it regular} time intervals we compute the
brightness profile along the disc midplane assuming a perfectly
edge-on disc. At each projected  distance from the star and
because the disc is assumed to be optically thin, the surface
brightness is obtained by integrating along the line
of sight the flux scattered by the dust particles.
The stellar flux received by a grain is assumed to drop
with the square of the distance from the star
(stellar flux dilution in an optically thin environment).
Scattering is assumed to be gray and isotropic. Gray scattering
implies that the scattering cross section is proportional
to the geometric cross section. For the isotropic assumption, we refer to
the discussion in \citet{aug01} who showed that the anisotropic
scattering properties of the grains only affect the regions below
$\simeq 80$\,AU as far as the scattered light radial
brightness profile along the disc midplane is concerned.
The final profile is progressively obtained by adding
these instantaneous profiles. This procedure is stopped
when no further significant evolution
of the total profile is observed at projected distances smaller
than $500\,$AU, either because particles eventually escape the
system or are being collisionally destroyed.

Our procedure is implicitly equivalent to assuming an
arbitrarily constant dust production rate from parent bodies.
The absolute value of the "real" production rate
is here an unconstrained parameter that does not affect
the results since we are only concerned with the shape
of the surface brightness profile and not its absolute intensity.
In other terms, only the relative size distribution of the produced dust matters
in our approach and not the exact amount of dust produced by
collisions. Our results might be rescaled to any absolute flux level
through the $C_{0}$ coefficient
because the collisional lifetime of dust in the 
models has been scaled to the observed value rather than being dependent 
on the density of material in the disc.

\section{Results}

For sake of clarity, we will focus on three observational
parameters, namely the two flux ratios in scattered light along the disc
midplane $F_{40/100}=F_{40\ma{AU}}/F_{100\ma{AU}}$ and
$F_{400/100}=F_{400\ma{AU}}/F_{100\ma{AU}}$ and the averaged
{power law index $p\dma{out}$ of the radial distribution profile
measured between $150$ and $400\,$AU. The reference observational
values deduced from \citet{heap00} are $F_{40/100} \simeq 0.3$,
$F_{400/100} \simeq 5\,10^{-3}$ (by extrapolating their profile)
and $-5.5<p\dma{out}<-4.5$.

\subsection{No gas}

\begin{figure}[tbp!]
\includegraphics[angle=0,origin=br,width=\columnwidth]{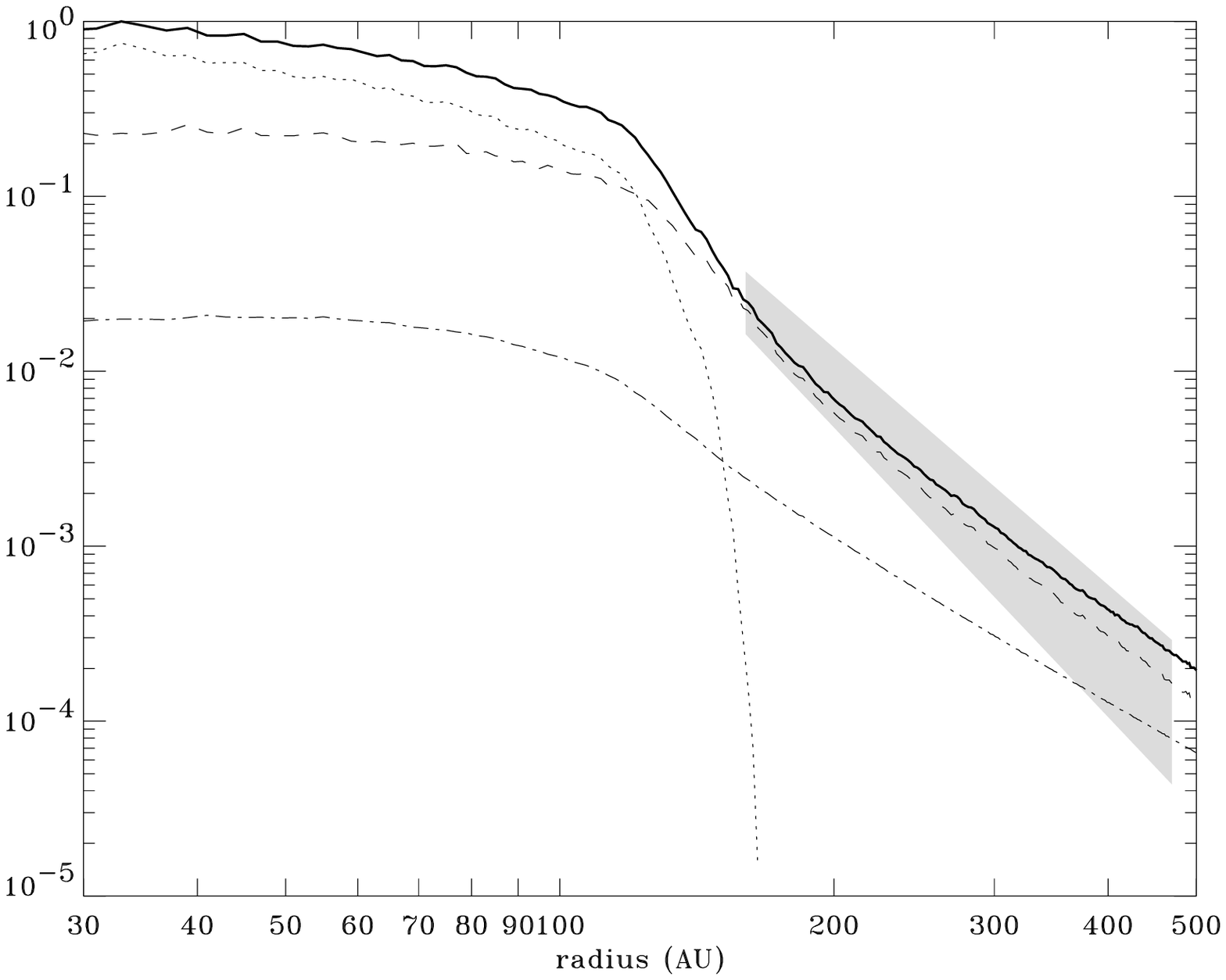} 
\caption[]{Simulated scattered light mid--plane profile in a gas free
disc (solid line).
The flux is arbitrary rescaled so that its maximum value equals 1.
The dotted line shows the contribution of the $0.005<\beta<0.05$ population,
the dashed line that of the $0.05<\beta<0.5$ and the dash-dot line that
of the $0.5<\beta<25$. The grey area shows the range of plausible
profiles in the outer region, as deduced from \citet{heap00} observations
for the SE and NW sides. It corresponds to the area bordered by $r^{-4.5}$
and $r^{-5.5}$ slopes derived from the flux value at 150\,AU.
} 
\label{nogas} 
\end{figure} 

We first perform an academic run with no gas drag
in order to provide us with a reference case to which
the following runs might be compared.

The synthetic scattered light profile displayed in Fig.\,\ref{nogas}
is close to the one obtained by \citet{aug01}, with
$F_{40/100}\simeq 0.35$ and $F_{400/100}\simeq 5.10^{-3}$.
However, we find $p\dma{out}\simeq -4.7$ while \citet{aug01} got
$p\dma{out}$ between $-5$ and $-5.5$. This small difference has two causes.
First of all, the presence of high $\beta$ ($>0.5$) grains ignored
in the dynamical approach of \citet{aug01} moderately contributes to the flux
in the outer regions (assuming a distribution for the $\beta>0.5$
particles similar to those with $\beta\simeq 0.5$, \citet{aug01} also
found a similar trend). Secondly, we consider here
the collisional lifetime of the particles whereas \citet{aug01}
did a phase mixing of randomly generated particle orbits. The
dependency of $t\dma{coll}$ with distance to the star tends to
increase the density of the $\beta \simeq 0.5$ particles which spend
a significant amount of time in the outer regions, hence the slightly
flatter slope beyond 150\,AU. However, this $-4.7$ slope is still in
agreement with the radial power-law index measured by \citet{heap00}
from the HST/STIS scattered light images: about $-4.5$ for
the northeast side of the disc and  about $-5.5$ for the southwest side.
It should be noted that a slope index close to -5
for the scattered profile beyond the parent body region is
in agreement with the results of \citet{lec96} who proved this to be a 
natural tendency for such systems.

\subsection{High gas density}

\begin{figure} 
\includegraphics[angle=0,origin=br,width=\columnwidth]{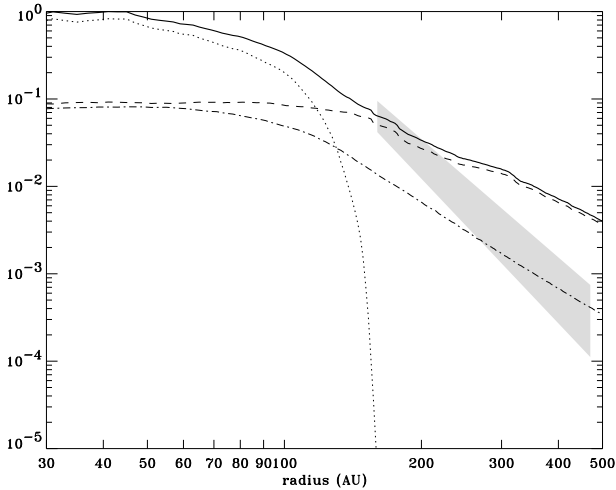} 
\caption[]{Same as Fig.\ref{nogas} but for the high gas density
disc (case 1 in Table\,\ref{tabinit}).
} 
\label{highgas} 
\end{figure} 

In this section we consider the high gas density case,
with $n_0=10^{6}$cm$^{-3}$
(case 1 in Tab.\,\ref{tabinit}), as inferred
by \citet{thi01} and as required by \citet{brand04} in order to
have hydrogen to act as a ``braking agent'' for observed high-$\beta$
gaseous species.

Differences with the gas free case remain negligible in the
inner 100\,AU region (Fig.\ref{highgas}), where
the flux is dominated by the biggest particles with 
$\beta<0.05$
which are only moderately affected by gas drag (see Fig.\ref{bprofs}).
Furthermore, it should be noted that the limiting value
$\beta\dma{steady}$, separating the regime where gaseous friction
leads to inward drift ($\beta<\beta\dma{steady}$) from the
regime of outward drift ($\beta>\beta\dma{steady}$),
falls within this size-range (see Fig.\ref{bprofs}b).
Using Equ.\,27 and 28 of \citet{take01} with the present gas density, one gets
indeed $\beta\dma{steady}=0.034$.

In the $r>150\,$AU region, however, differences with the previous
case are striking. The obtained brightness profile is clearly
incompatible with the \citet{heap00} data, with
$F_{400/100}\simeq 2.10^{-2}$ and $p\dma{out}\simeq -2.7$.
This is mainly due to the strong increase of the
$0.05<\beta<0.5$ particles contribution.
This population is strongly affected by gaseous friction
and is progressively driven into the
outer disc, hence flattening the radial profile,
whereas in the gas free case only particles very close to the
$\beta = 0.5$ limit have orbital eccentricities high enough to reach
these regions (Fig.\,\ref{bprofs}).
This outward drift has the additional effect
of lengthening the collisional lifetime of these particles
(because of the radial dependency of the collision rate, cf. Equ.\,\ref{tcoll}).
The contribution of the $\beta>0.5$ particles is also increased with
respect to the gas free case. For these particles the effect of gas drag is
on the contrary to slow down their radial escape, hence prolonging
their stay in the disc (Fig.\,\ref{bprofs}).
Their contribution to the total flux remains nevertheless marginal.
It is
one order of magnitude smaller than the flux produced by the
$0.05<\beta<0.5$ particles between 100 and 400\,AU (Fig.\,\ref{highgas}).

\subsection{Exploring the gas density}

\begin{figure} 
\includegraphics[angle=0,origin=br,width=\columnwidth]{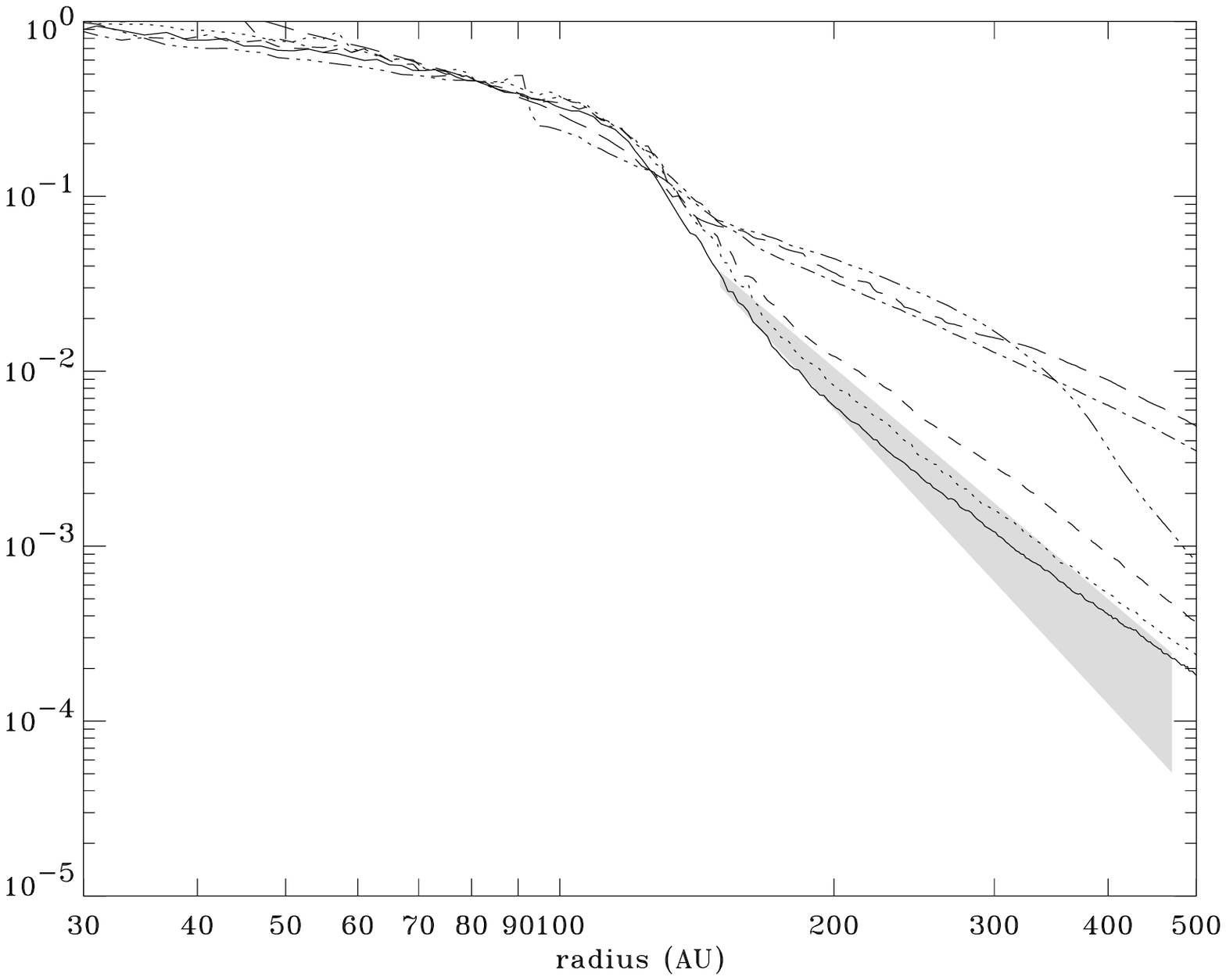} 
\caption[]{Global profile for different gas disc prescriptions.
Solid line: gas free profile of Fig.\,\ref{nogas}.
Then, from bottom to top of the profile values at 500\,AU:
Solar abundance reconstructed gas disc, i.e.
\citet{brand04} profile with $n_{0}=2.25\,10^{3}$cm$^{-3}$,
same profile with $n_{0}=10^{4}$cm$^{-3}$,
\citet{brand04} profile with $n_{0}=10^{6}$cm$^{-3}$
truncated at 150\,AU,
\citet{brand04} profile with $n_{0}=10^{5}$cm$^{-3}$,
Hayashi profile with a total disc mass of $40\,M_{\oplus}$.
These models correspond respectively to cases 2, 3, 5, 4
and 6 in Table\,\ref{tabinit}.} 
\label{diffgas} 
\end{figure} 

Taking the low gas density distribution  reconstructed by \citet{brand04}
assuming solar abundances leads to a luminosity profile which
is almost undistinguishable from the gas free one (Fig.\,\ref{diffgas}).
Additional runs assuming the profile of Equ.\,\ref{sod} with varying
$n_{0}$ show that departures from the gas free case, and
thus incompatibility with the observed \citet{heap00} luminosities, become
significant for $n_{0}>10^{4}$cm$^{-3}$. 
We also performed an academic run with the high gas density distribution
truncated beyond 150\,AU. But this leads to a luminosity profile much too
high around 150--300\,AU, mainly because of the accumulation
of the $0.05<\beta<0.5$ grains in this region. This is an effect already
observed by \citet{take01} for small grains in truncated gaseous discs.
An additional run with a total gas mass equal to that of case
2 ($40\,M_{\oplus}$) but assuming a solar type profile in $r^{-2.75}$
\citep{haya81} also leads to strong departures from the observed 
luminosity profile in the outer region.

Note also that the dust luminosity profile in the inner 100\,AU region
is almost independent from gas drag effects, at least for the density
range explored here.

\subsection{Exploring the dust size distribution in the high gas density case}

\begin{figure} 
\includegraphics[angle=0,origin=br,width=\columnwidth]{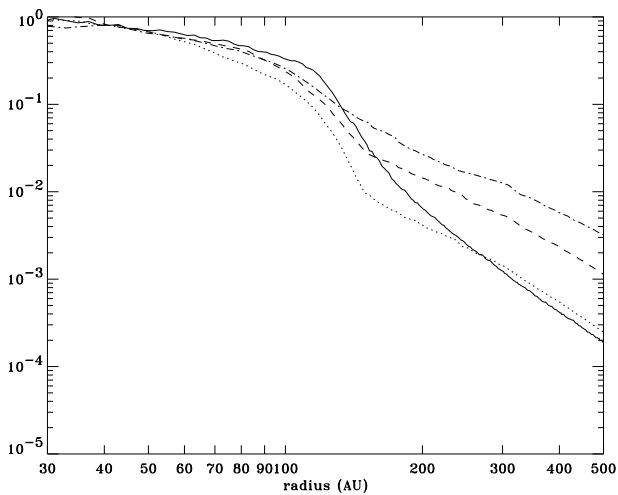} 
\caption[]{Profiles obtained for different dust size distributions
in the high gas density disc (case 1).
dash-dot line: $\d N \propto s^{-3.5}\d s$, dashed line: $\d N \propto
s^{-3}\d s$, dotted line: $\d N \propto s^{-2.5}\d s$. The solid line
is the reference gas free profile of Fig.\,\ref{nogas}
} 
\label{dustsiz} 
\end{figure} 

All luminosity profiles sofar were computed assuming that dust particles
follow the classical $\d N \propto s^{-3.5}\d s$ size distribution, which is
the prescription usually taken in similar studies
\citep[e.g.][]{aug01}. However, it should be stressed that ``real''
size distributions may significantly depart from this theoretical
collisional equilibrium power law, especially
for particles close to the blow-out size (the reader might refer
to \citet{the03} for a detailed discussion on this issue).

Accurately estimating the size distribution exceeds the scope of
the present work, but we have tested alternative size-distributions
power laws in order to explore how the final profile depends on this
parameter. 
Figure\,\ref{dustsiz} clearly shows that reducing the power-law index of
the size distribution indeed leads to a drop of flux beyond
150\,AU. However, the price to pay for this better fit in the outer
disc is a significant and troublesome flux reduction in the 100-150\,AU
region. This leads to profiles which are as bad a fit to observational data
as in the $\d N \propto s^{-3.5}\d s$ case.
This flux reduction could in principle be compensated by allowing the radial
distribution of the parent bodies to extend further out. However, this would in
turn lead to an increase of the flux beyond 150\,AU, and yet departure
from the good match originally obtained there,
because the luminosity
profile in these outer regions is mainly due to particles which originate from
the highest parent body density region.

\section{Discussion}

The above results all tend to point towards an incompatibility between
high gas densities and dust luminosity profiles matching observations.
One could wonder however if these discrepancies could not be eliminated
by assuming parent bodies distributions that differ from that
proposed by \citet{aug01} in the context of a {\it gas free} system.
In this case the problem would get hopelessly degenerated: for every
gaseous disc considered there would always be an ad-hoc parent body
distribution producing a good fit to the observations.

We do not believe this to be the case because the simulations
show that parent body distributions cannot be arbitrarily chosen
neither in the inner nor in the outer regions of the disc. 
In the inner ($<100$\,AU) region, the luminosity profile
is dominated by big grains with  $\beta<0.05$ (see Fig.\,\ref{highgas}).
These particles are only
weakly affected by gas drag and remain on nearly circular orbits matching
those of the parent bodies producing them. It follows that,
for the inner disc, a change
of parent body distribution would automatically lead to departures from the 
best-fit luminosity profile which cannot be compensated by gaseous friction
effects, at least within the gas density range explored in this paper
(additional test runs have shown that significant departures from the
gas free profile are found, in the inner disc, when $n_{0}>10^{7}$cm$^{-3}$,
which corresponds to unrealistic large amounts of gas).

Beyond 150\,AU the problem gets more complex and there might in principle
be an alternative parent body distribution able to ``bend'' the outer
disc profile towards the ``right'' one even for the high gas density case.
However, it is easy to see that in order to achieve this,
this alternative distribution should
act as to $decrease$ the outer disc luminosity by more than an order
of magnitude (Fig.\,\ref{highgas}). 
This cannot be done by adjusting the parent body distribution in the
outer disc, since in the reference case this distribution is already truncated
beyond 150\,AU, so any alternative distribution in the outer disc
can only add objects there and
result in an
$increase$ of the luminosity.
Thus in order to lower the luminosity profile in the outer disc
the only possibility 
is to change the parent body distribution in the $inner$ disc.
But, as has just been shown,
this cannot be done without in turn strongly modifying the luminosity
profile in the inner regions (which would result in departing from
a good fit of observations), thus solving one problem by creating another.
As a consequence, we do believe the \citet{aug01} parent body profile
to be a rather robust assumption.

Alternatively, one may consider different dust size distributions,
for instance a shallower distribution in order to lower the flux in the
outer regions where small grains dominate. However, Fig.\,\ref{dustsiz} shows that
this doesn't work either. We cannot completely exclude
the possibility that a fine-tuned fit might be found by adjusting
the size distribution at every distance from the star, but
such a thorough
parameter exploration is beyond the scope of the present work. However, even if
such an ad-hoc fit should be found, it would have the disadvantage 
of being less generic.

Another parameter space that was left unexplored here is the grain
scattering properties. However, for dust radial distributions relatively similar
to the present ones, \citet{aug01} has shown that assuming anisotropic
scattering properties for the grains would not significantly affect
the flux profile beyond $\sim$80\,AU, {\it i.e.} the region of interest
for the present discussion.

Under these assumptions we thus do believe that the present simulations rule
out the presence of large amounts of gas around \bp, especially in the outer
parts of the disc. Assuming the profile of \citet{brand04}, the upper limit
on the $n_{0}$ value is $\simeq 10^{4}$cm$^{-3}$ corresponding to a total
gas mass of $\simeq 0.4\,M_{\oplus}$. This corresponds to a gas to dust ratio
smaller or approximatively equal to 1 at most.

\section{Summary and conclusion}

We have numerically investigated the dynamics of dust particles
suffering a
gas drag force in the edge-on \bp\ disc. 
From our simulations we were able to compare theoretical
scattered light profiles along the disc midplane
to the reference images obtained by \citet{heap00}.
Several gas disc prescriptions and a broad range of grain sizes (from
$\beta=0.005$ to $\beta = 25$) have been explored. We assumed
that dust particles are produced by parent bodies following
the distribution derived by \citet{aug01}, an assumptiom which our
simulations prove to be relatively robust.

Our main result is that high density gas discs 
always lead to dust distributions whose luminosity profiles
are strongly incompatible with the observations in the outer
regions of the disc (beyond 150\,AU), while the surface brightness profile
in the inner regions is poorly sensitive to the effect of
gas drag on the grains.
Assuming the gas density profile of \citet{brand04}, gas discs with total
mass above $\sim\, 0.4\,M_{\oplus}$ seem to be ruled out. This value is
consistent with the H$_2$ upper limit obtained by \citet{lec01}
and with the total gas mass estimated by \citet{kamp03}. It is also consistent
with the gas mass upper limit found by \citet{che04} based
on recent Spitzer mid-infrared observations. Unless the gas disc
is surprisingly very extended so that the dust dynamics is not affected
by gas drag within the first hundreds of AU from the star, our results
tend to rule out the much higher gas mass estimate ($50 M_{\oplus}$)
derived by \citet{thi01}. However, this leaves
open the puzzling problem of high $\beta$ ions observed on bound orbits,
since it also rules out the gaseous braking agent hypothesis for which at
least $40 M_{\oplus}$ of hydrogen are required according to \citet{brand04}.
Our results suggest that other scenarios, like braking by other
chemical species or by magnetic field, have definitely to be investigated.
This may lead to a better estimate of the mass of gaseous species
insensitive to stellar radiation pressure required to stop high $\beta$ ions
\citep[][]{beu05}.
The approach developped in the present paper could also in the future
be applied to similar debris disc systems, in particular the
recently discovered disc around \mic, which 
is coeval with \bp\ \citep{kalas04}.

\begin{acknowledgements}
The authors thank C. Chen and K. Stapelfeldt for providing informations
regarding the interpretation of their Spitzer observations of \bp.
This work is supported by the European Research Training Network
"The Origin of Planetary Systems" (PLANETS, contract number
HPRN-CT-2002-00308).
\end{acknowledgements}

{}
\clearpage

\end{document}